\documentclass{article}
\usepackage{spconf,amsmath,epsfig}

\let\OLDthebibliography\thebibliography
\renewcommand\thebibliography[1]{
  \OLDthebibliography{#1}
  \setlength{\parskip}{0pt}
  \setlength{\itemsep}{0pt plus 0.3ex}
}

\usepackage{xcolor}

\usepackage{times}
\usepackage{soul}
\usepackage{url}  
   
\usepackage[hidelinks]{hyperref}
\usepackage[utf8]{inputenc}
\usepackage[small]{caption}
\usepackage{graphicx}
\usepackage{amsmath}
\usepackage{amsthm}
\usepackage{booktabs}
\usepackage{algorithm}
\usepackage{algorithmic}
\usepackage[para,online,flushleft]{threeparttable}
\usepackage{makecell}
\usepackage{booktabs}
\usepackage{multirow}
\usepackage{algorithm} 
\usepackage[algo2e]{algorithm2e} 
\usepackage{hyperref}
\usepackage[hyphenbreaks]{breakurl}
\usepackage{url}

\newcommand{\M}{FAIVConf}

\begin{document}\sloppy

\def\x{{\mathbf x}}
\def\L{{\cal L}}

\title{FAIVConf: Face enhancement for AI-based Video Conference with Low Bit-rate}
%
\name{Zhengang Li$^{1,2}$, Sheng Lin$^{1}$, Shan Liu$^{1}$, Songnan Li$^{1}$, Xue Lin$^{2}$, Wei Wang$^{1}$ and Wei Jiang$^{1}$\thanks{Work done when Zhengang Li interned at Tencent Media Lab.}}
\address{$^{1}$Tencent Media Lab, $^{2}$Northeastern University\\
$^{1}$\{barneylin, shanl, sunnysnli, rickwang101, vwjiang\}@tencent.com, @gmail.com;\\$^{2}$\{li.zhen, xue.lin\}@northeastern.edu.}
\maketitle

\begin{abstract}
Recently, high-quality video conferencing with fewer transmission bits becomes a very hot and challenging problem. 
We propose {\M}, a specially designed video compression framework for video conferencing, based on the effective neural human face generation techniques.
{\M} brings together several designs to improve the system robustness in real video conference scenarios: face swapping to avoid artifacts in background animation;
facial blurring to decrease transmission bit-rate and maintain quality of extracted facial landmarks; and dynamic source update for face view interpolation to accommodate a large range of head poses.  
Our method achieves significant bit-rate reduction in video conference and gives much better visual quality under the same bit-rate compared with H.264 and H.265 coding schemes.
\end{abstract}

\section{Introduction}

There has been a growing demand for video conferencing solutions to provide stable and high-quality video communication with low latency. Due to the limited bandwidth resources, it is critical to effectively reduce the video bit-rate to ensure a smooth user experience with little transmission lag or artifacts. Traditional video coding tools such as H.264/H.265 aim at compressing general video content, which, albeit may be improved through continuous study, are suboptimal for the video conferencing scenario with its unique characteristics. 

Human faces are especially important in video conferences, which usually occupy a large portion of the frames and also are the main focus of the frames. Therefore, it is particularly important to improve the compression quality of the human faces in this scenario. With the great success of the generative adversarial network (GAN), significant progress has been made in human face generation. These methods \cite{nirkin2019fsgan,siarohin2019first} usually generate face images based upon face-related semantic information such as facial landmarks~\cite{wang2020deep}, estimated poses~\cite{ruiz2018fine}, and segmentation masks~\cite{ronneberger2015u}. Such face-related data have a much smaller scale compared to the original frames, which makes it possible to build an extremely low bit-rate and high-quality coding framework for enhancing visual quality in video conferencing.
For example, NVIDIA's video conferencing solution \cite{wang2021one} transfers only a keypoint representation of faces, which is used to reconstruct the source frames in decoder. Although theoretically appealing, such methods suffer from several challenges in real applications, such as (complex) background generation, large differences for pose transfer, large illumination mismatch, facial occlusions, \textit{etc.}

Targeting at robust performance in real applications, we propose {\M}, a video compression framework specially designed for practical video conferences. 
Our contribution can be summarized as follows:

First, we propose a simple yet effective facial blurring mechanism to decrease the bit-rate of transmitting the facial area while keeping the essence of facial features, since blurring can effectively reduce the blocky artifacts caused by heavy compression, which affects accuracy of facial landmark extraction. 
To ensure high-quality face generation with a large range of head poses in real applications, we propose a dynamic updating mechanism for the view interpolation method in the face reenactment module. Based on the above improvements, we propose the {\M} to reduce bit-rate and optimize the video transmission quality in the video conference scenario.
Our proposed framework achieves significant bit-rate reduction, \textit{i.e.}, only 0.001875 bits per pixel for transmitting a conference video with 800$\times$800 pixels, which is about 0.8\% of streaming the original video. Compared with the commercial H.264 and H.265, our method gives much better visual quality under the same bit-rate.

\section{Motivation}

\label{sec:motivation}
GAN-based guided face generation enables highly efficient compression in the scenario of video conference.
Guided by only facial landmarks, one can achieve optimal compression rate since only one source image and successive keypoints need to be  transferred to the decoder for face generation. 
Such methods assumes simple clean background and limited pose difference between the source and the drive image, without significant occlusions over the face area. 
Unfortunately, these assumptions are often violated in practical video conference sessions, and these methods suffer from severe artifacts.

\begin{figure}[t]    
 \centering
\includegraphics[width=0.75\columnwidth]{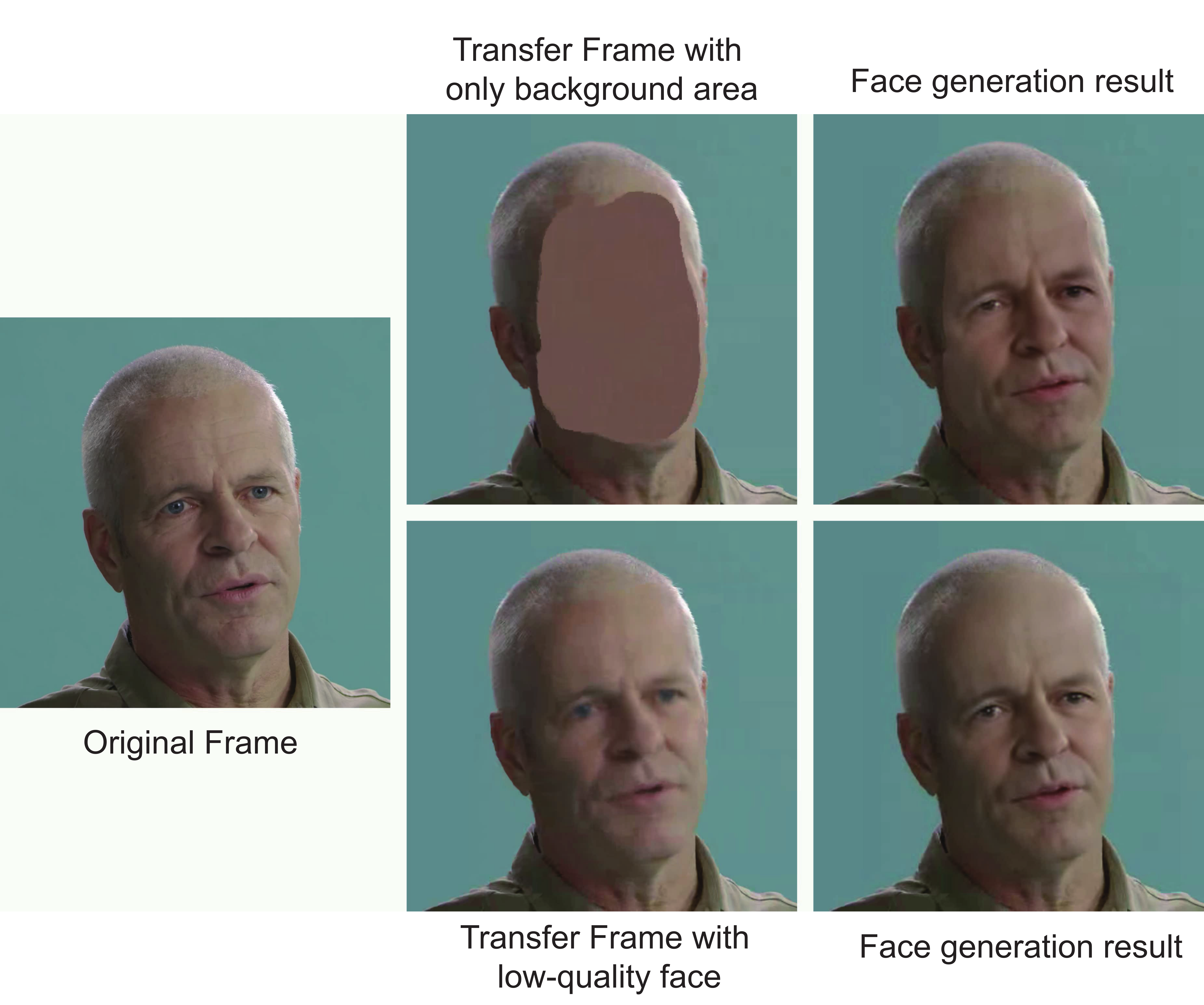} \\
\caption{Example comparison of face generation by: 1. only using the background information; 2. using both background information and a low-quality face of the original frame.}
\vspace{-0.5cm}
\label{fig:gaussianblur}
\end{figure}

To avoid modeling unconstrained backgrounds, we need to confine the face generation only to the true face area, and compress facial and non-facial areas separately. To avoid the difficulty of face reenactment with large pose and illumination differences, we also transfer a low-quality (\textit{e.g.}, highly blurred) version of the original frame to guide the face generation. 
Fig.~\ref{fig:gaussianblur} gives an example of face generation with or without using the low-quality face area in the original frame. The face generation result with Gaussian blurred face (bottom right) can simulate the lighting condition much better than the result using only background (top right).
In addition, we propose dynamic updating face view interpolation mechanism using multiple source images to further improve the robustness of the generated face with regard to flexible head pose in real video conference.  
One caveat here is the increased transmission overhead. The background is compressed and transferred by itself to maintain the fidelity of the video background. 
Since face is the main changing part in video conferences, which usually occupy the majority of bits. The background can be compressed efficiently in general. 
Regarding the detected face area, to reduce the transmission overhead as much as possible, we perform facial landmark extraction and face segmentation all on the decoder side, and only transfer a blurred version of an extended face area. 
We conduct Gaussian blur before encoding the face area to achieve a good balance between reduced bit-rate and quality of extracted landmarks. We observe that the accuracy of facial landmark extraction is quite sensitive to the blocky artifacts from compression. Gaussian blur can not only perform low-pass filtering to reduce the 
blocky artifacts for high compression ratio, but also lead to better bit-rate due to the blurred content. 

\section{The \M~Framework}

\begin{figure*}[tb]    
 \centering
\includegraphics[width=2.0\columnwidth]{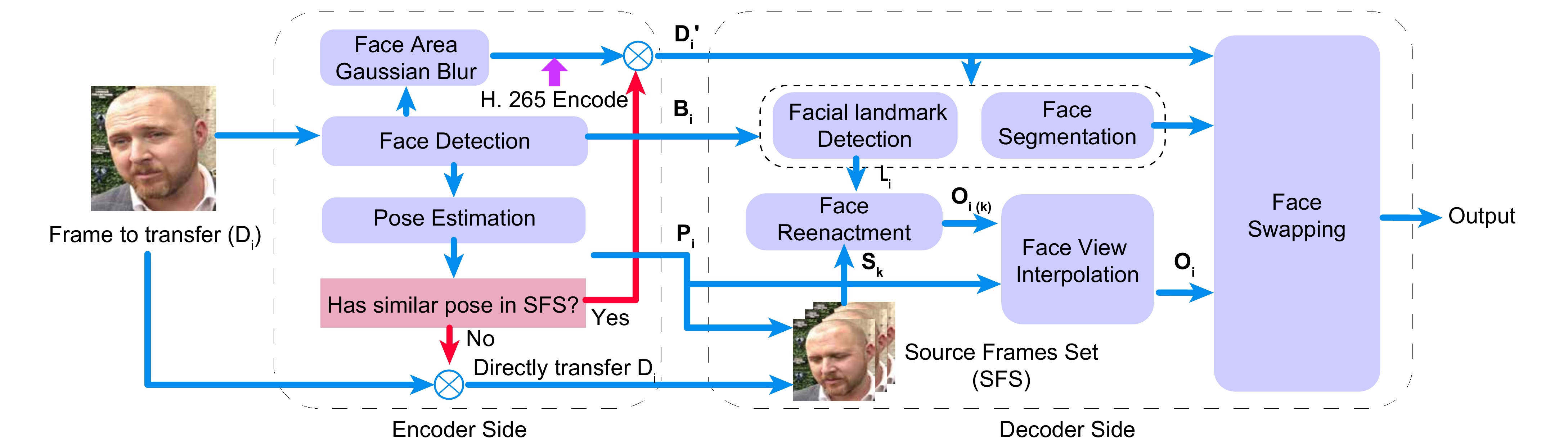} \\
\caption{The overall architecture of our proposed {\M} framework which is separated into encoding and decoder sides. The blue arrows indicate the data flow in the framework, and the red arrows indicate the control signals for the data flow.} \label{fig: distribution}
\vspace{-0.5cm}
    \label{fig:framework}
\end{figure*}

As shown in Fig.~\ref{fig:framework}, the proposed {\M} framework mainly consists of two parts. \textbf{Encoder} and \textbf{Decoder}.

\noindent{\bf Encoding Process.}
Let $\{D_1, D_2,D_3...D_i...D_n\}$ be the input conference video frames (driving video) to be transferred, where $D_i$ is the $i$-th driving frame in a total number of $n$ frames. Let $\{S_1, S_2, S_3...S_m\}$ be a set of source frames which has been collected from encoder side and stored in decoder side.
The goal of encoding process is to extract the feature of each  frame which can be transferred with small bandwidth to decoder. The transferred feature should be enough for the decoder to generate the output video $\{O_1, O_2, O_3...O_n\}$, which is highly similar to driving video. 

Specifically, we first apply face detection on the driving frame to get the bounding box of the face position $B_i$. Then we compute the head pose $P_i$ and perform face segmentation in parallel. The head pose information is compared with that of each source frame to see if similar poses exist. 
If the answer is yes, Gaussian blur will be applied on the segmented face area to decrease the video compression load. Then, the blurred frame will be encoded by the H.265 codec, and transferred to decoder side with the face bounding box and head pose. If the answer is no, the whole frame $D_i$ will be transferred to the decoder side directly, and the face area in $B_i$ will be stored in decoder as a new source face. 

\noindent{\bf Decoding Process.}
The decoding process is executed at the user terminal. After receiving the transferred data of the $i$-th driving frame, \textit{i.e.}, the recovered frame $D^{\prime}_i$ (after H.265 decoding), the head pose $P_i$ (expressed as Euler angles), and the face bounding box $B_i$, the decoder extracts both the facial landmarks $L_i$ and the segmented face of $D^{\prime}_i$ in parallel.
Then one or several source frames will be chosen to perform the face reenactment according to the head pose similarity as shown in Sec.~\ref{Face View Interpolation}. The landmarks $L_i$ and reenacted output from the $k$-th source frame is represented as $O_{i(k)}$ ($1 \leq k \leq m$): 

\begin{equation}
\begin{aligned}
L_i = G_l(B_i; D^{ \prime}_i); \\
O_{i(k)}=G_r(S_k; L_i)
\end{aligned}
\end{equation}
where $G_l$ and $G_r$ are the facial landmark detector and reenactment generator, respectively. Then we compute the interpolated reenacted output $O_i$ through face view interpolation, which will be discussed in detail in Sec.~\ref{Face View Interpolation}. Then,  inpainting~\cite{yeh2017semantic} and blending~\cite{wu2019gp} will be performed to swap $O_i$ back into the transferred driving frame $D^{\prime}_i$ within the bounding box area defined by $B_i$ to generate the final output frame.

\noindent{\bf Dynamic source pool Update for Face View Interpolation.}
\label{Face View Interpolation}
To overcome the difficulty of image animation posed by a large range of facial orientations in real applications, we incorporate the face view interpolation technique~\cite{nirkin2019fsgan} in our framework. This method allows using multiple source frames rather than a single image to generate the face reenactment result and makes it possible to synthesize faces with different pose with little distortion. We further improve this method by dynamically updating the source frames through the video conference session to ensure small pose differences for quality reenactment.

For the source frames set (SFS) $\{S_1, S_2,...S_m\}$, the corresponding Euler angles are  represented as $\{e_1, e_2,...e_m\}$. Each angle $e_j$ consists of yaw $y_j$, pitch $p_j$ and roll $r_j$.
We first project the Euler angle into a plane by dropping roll, \textit{i.e.}, $\{e_{1(y, p)}, e_{2(y, p)},...e_{m(y, p)}\}$, and then remove points in the angular domain that are too close to each other with a distance threshold $\mathcal{L}$. A mesh can be built based upon the remaining points in the angular domain by Delaunay Triangulation.

For the $i$-th driving frame with head pose $P_i$, its corresponding projected point in the angular domain is $e^d_{i(y, p)}$. We find the most related three source frames $S_{k_1}$, $S_{k_2}$, $S_{k_3}$ at the vertices of the triangle which contains $e^d_{i(y, p)}$. With the barycentric coordinates $(\lambda_{k_1}, \lambda_{k_2}, \lambda_{k_3})$ of $e^d_{i(y, p)}$ in the triangle, the face view interpolation result $O_{i}$ can be calculated as:
\begin{equation}
O_{i}=\sum\nolimits_{r=1}^{3} \lambda_{k_r} O_{i(k_r)},
\end{equation}
where $O_{i(k_r)}$ is the reenacted output from the source frame $S_{k_r}$ corresponding to the $k_r$-th vertex.

In addition, we update the set of source frames dynamically. When the projected point of a driving frame in the angular domain is not in any of the existing triangles, we only use the source frame of the closest existing vertex, and set $\lambda$ as $1$. If a driving frame is far from all source frames in the angular domain, \textit{i.e.}, distances between the projected points of the driving frame and the source frames are all above a threshold $\mathcal{L}$, this driving frame will be added into the set of source frames and transferred to the decoder. The Delaunay Triangulation mesh will be rebuilt to include the new source frame to enable a more comprehensive range of head pose. 

\section{Experiments}

\noindent{\bf Experiment Setup.}
We used the mixed video sequences of the IJB-C~\cite{maze2018iarpa} and Human-centric video matting~\cite{humancentricmatting} dataset to train our reenactment model, and used the WFLW dataset~\cite{wu2018look} to train the facial landmark detector.
Hopenet~\cite{ruiz2018fine} was used as our pose estimator in our framework.

For the reenactment model training, we used two VGG-19 models~\cite{simonyan2014very} pretrained on VGGFace2~\cite{cao2018vggface2} and CelebA~\cite{liu2018large} datasets to compute the perceptual loss~\cite{johnson2016perceptual} towards face recognition and face attribute classification, respectively. This perceptual loss has been wildly used recently~\cite{nirkin2019fsgan,wang2021one}.  Besides that, we also used the multi-scale GAN loss~\cite{wang2019few} and the pixelwise loss in the training process to enhance the generator's performance.

\noindent{\bf Landmark Detection of Blurred Face.}

We first preprocessed the WFLW dataset with Gaussian blur within the face area to simulate the face blurring of frames transmitted between the encoder and decoder. 
Then, we trained HRNetV2 to extract facial landmark with the same setting in the work of~\cite{wang2020deep}. 
As shown in Table~\ref{tab:landmark}, we compare the normalized mean error (NME) of HRNetV2 (Trained on original WLFW) and HRNetV2-blur (trained on blurred WLFW data). 
We can see that training landmark detector with blurred data can significantly decrease the NME of all the subsets in the Gaussian blur situation, and the loss is just a little higher than that of the original model on non-blurred dataset. This means that HRNetV2-blur on blurred image has the same accuracy level compared with the original model on the same image without blurring. Therefore, HRNetV2-blur is good enough to handle transmitted frames in decoder.

\begin{table}
\centering
\caption{Facial landmark detection results (NME) on original WFLW dataset and the preprocessed WFLW dataset (Gaussian blur) with test and 5 subsets: large pose, expression, illumination, makeup, and occlusion. Lower is better.}
\scalebox{0.55}{
\begin{tabular}{llrrrrrrr}
\toprule
Model  & Dataset & Test & Large-Pose & Expression & Illumination & Makeup & Occlusion \\
\midrule
HRNetV2~\cite{wang2020deep} & WLFW     & 4.60  & 7.94  & 4.85  & 4.55  & 4.29  & 5.44 \\
HRNetV2~\cite{wang2020deep} & WFLW-blur& 13.22 & 27.15 & 11.81 & 13.50 & 12.33 & 16.24 \\
HRNetV2-blur  & WFLW                   & 5.26  & 8.87  & 5.63  & 5.11  & 5.12  & 6.20  \\
HRNetV2-blur  & WFLW-blur              & \textbf{5.76}  & \textbf{10.01} & \textbf{6.03}  & \textbf{5.84}  & \textbf{5.77}  & \textbf{7.07}  \\
\bottomrule
\end{tabular}
}
\vspace{-0.5cm}
\label{tab:landmark}
\end{table}

\begin{figure}[t]    
 \centering
\includegraphics[width=1.0\columnwidth]{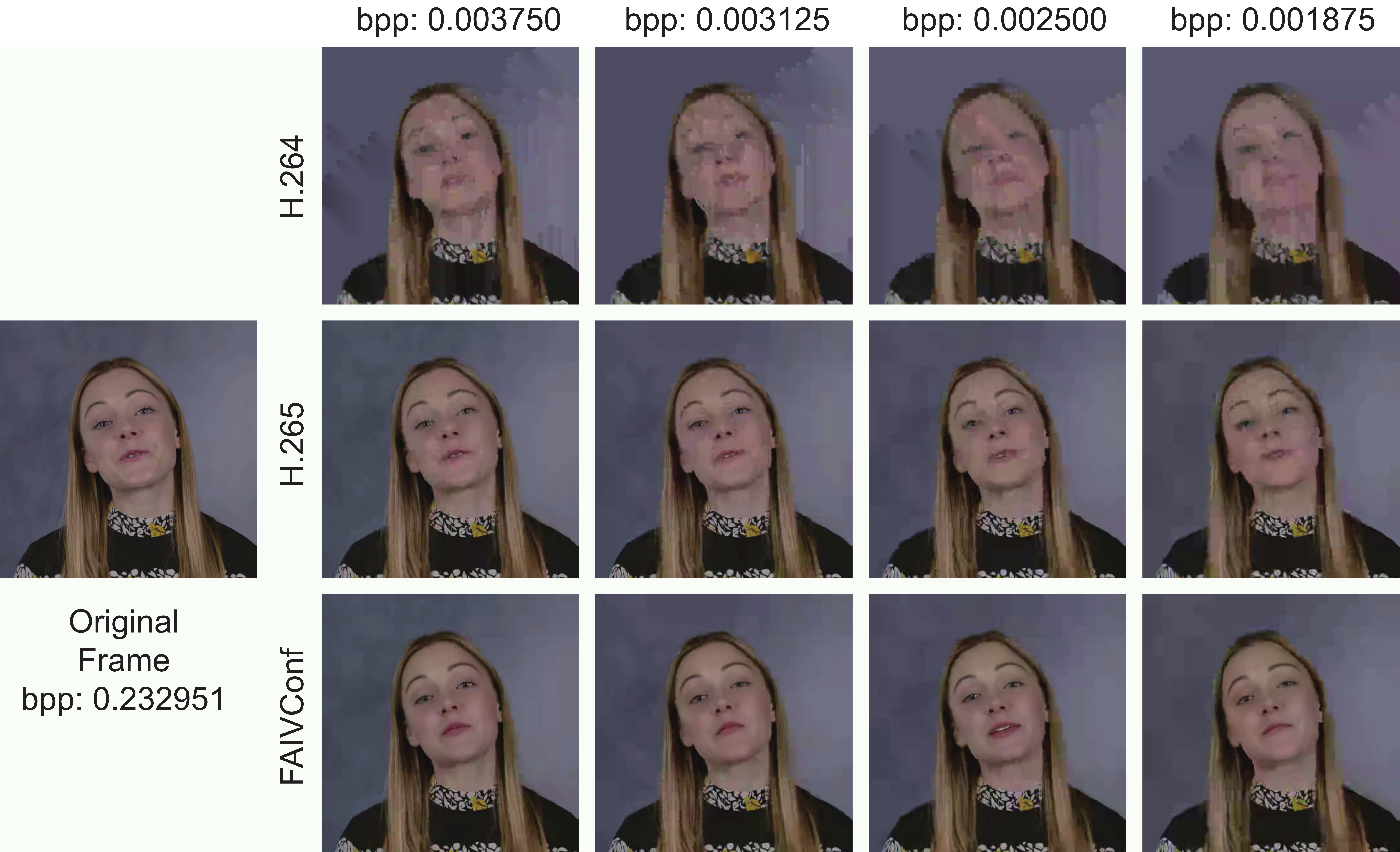} \\

\caption{Comparison of result at the same bit-rate with H.264/H. 265.} \label{fig: distribution}
 \vspace{-0.5cm}
\label{fig:H264}
\end{figure}

\noindent{\bf Performance on Conference Video.}
To evaluate our performance in video conference, we propose a High-resolution Talking Head (HSTH) dataset, which contains a scene where a single person is talking with high resolution up to 800$\times$800 collected from YouTube.
Our method can re-create a talking-head video at decoder side based on the transmitted encoded frame. Fig~\ref{fig:H264} represents the comparison between final result of our approach with that of using commercial H.264/H.265 coding scheme directly. To ensure fair comparison, we encoded the intermediate transmitted data in~\M~ with the same total bit-rate, \textit{i.e.}, the total summed bit-rate of $D^{\prime}_i$, $P_i$, and $B_i$ was equal to that of the encoded frame in H.264/H.265 coding scheme.
The leftmost image is the original frame (driving frame) in encoder. On the right side, the three rows show the result of H.264/H.265 coding and~\M. From left to right, the bit-rate of the transferred data decreases one after another. We also list the bpp (bits per pixel) of each level.
Our approach achieves consistent good performance in different bit-rate levels, while H.264/H.265 shows larger compression artifact as bpp gets lower. \M~can generate acceptable output even the transmission is only 0.001875 bbp, which is 0.8\% of the original video. We test the PSNR and SSIM value on our proposed HSTH dataset. For bits per pixel with the level of 0.002, the average PSNR and SSIM achieve 26.3dB and 0.89, respectively.

\begin{figure}[t]    
 \centering
\includegraphics[width=1.0\columnwidth]{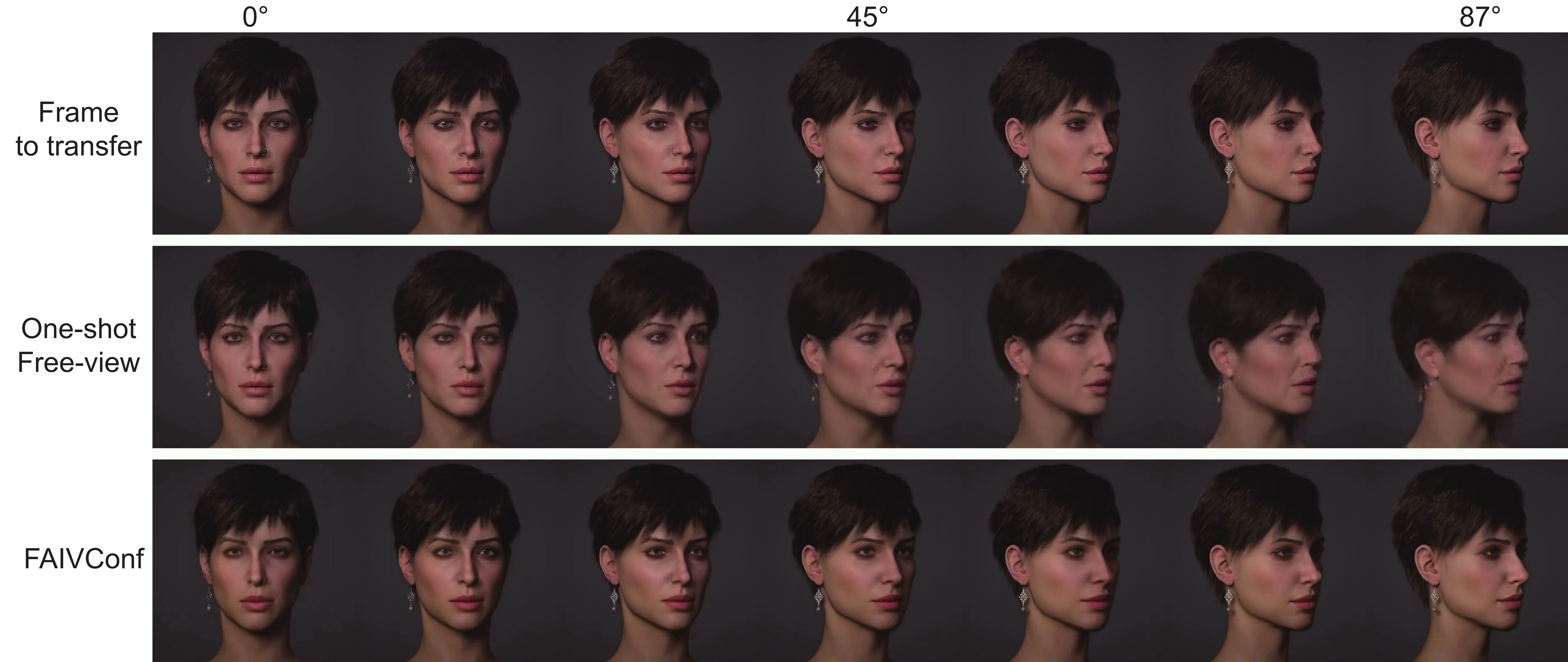} \\

\caption{Comparison of OSFV with~\M~under different head rotation angles. Our approach shows better performance especially when the angle is larger than $45^{\circ}$.} \label{fig:oneshot-vs-FAIVConf}
\vspace{-0.5cm}
\label{fig:angle}
\end{figure}

Fig~\ref{fig:angle} gives the comparison between OSFV (One-shot Free-view~\cite{wang2021one}) with our framework~\M. The first row shows the transferred video frames with the increasing head rotation angle from left to right. The second and third rows show the corresponding face generation results using OSFV and~\M, respectively. Here, the leftmost driving frame with $0^{\circ}$ rotation angle is used as the source image of OSFV. When the head rotation angle is small, both frameworks provide good results. However, as the angle increases, especially when it is greater than 45 degrees, the face generated by OSFV gradually deforms and becomes flattened, while~\M~can still generate high quality images which maintain the identity and fidelity similar to the original frames.

\section{Conclusion}

We introduce \M, a framework dedicated to highly efficient video compression for the video conference scenario. 
Compared with the commercial H.264/H.265 coding schemes, our approach achieves much better visual video quality with less transmission data. We achieve significant  bit-rate reduction under similar visual quality via only $0.8\%$ of streaming the original video in a resolution of 800$\times$800 pixels.

\bibliographystyle{IEEEbib}
\bibliography{icme2022template}

\end{document}